\documentstyle[12pt]{article}

\begin{document}

\title{{\bf {Vacuum quantum effect for curved boundaries in static
Robertson--Walker spacetime }}}
\author{ M. R. Setare \thanks{%
E-mail: rezakord@ipm.ir} \\
{Department of Science, Payame Noor University, Bijar, Iran} \\
J. Sadeghi\thanks{%
E-mail: pouriya@ipm.ir }\\
{\ Sciences Faculty, Department of Physics, Mazandaran University }\\
{\ P .O .Box 47415-416, Babolsar, Iran}}
\maketitle

\begin{abstract}
The energy-momentum tensor for a massless conformally coupled scalar field
in the region between two curved boundaries in $k=-1$ static
Robertson--Walker spacetime is investigated. We assume that the scalar field
satisfies the Dirichlet boundary condition on the boundaries. $k=-1$
Robertson--Walker space is conformally related to the Rindler space, as a
result we can obtain vacuum expectation values of energy-momentum tensor for
conformally invariant field in Robertson--Walker space from the
corresponding Rindler counterpart by the conformal transformation.
\end{abstract}



\newpage 

\section{Introduction}

The Casimir effect is regarded as one of the most striking manifestation of
vacuum fluctuations in quantum field theory. The presence of reflecting
boundaries alters the zero-point modes of a quantized field, and results in
the shifts in the vacuum expectation values of quantities quadratic in the
field, such as the energy density and stresses. In particular, vacuum forces
arise acting on constraining boundaries. The particular features of these
forces depend on the nature of the quantum field, the type of spacetime
manifold and its dimensionality, the boundary geometries and the specific
boundary conditions imposed on the field. Since the original work by Casimir
in 1948 \cite{Casi48} many theoretical and experimental works have been done
on this problem (see, e.g., \cite{Most97,Plun86,Lamo99,Bord01} and
references therein). \newline
The Casimir effect can be viewed as a polarization of vacuum by boundary
conditions. Another type of vacuum polarization arises in the case of an
external gravitational fields \cite{Birrell,Grib94}. Casimir stress for
parallel plates in the background of static domain wall in four and two
dimensions is calculated in \cite{{set1},{set2}}. Spherical bubbles immersed
in different de Sitter spaces in- and out-side are considered in \cite%
{{set3},{set4}}. \newline
It is well known that the vacuum state for an uniformly accelerated
observer, the Fulling--Rindler vacuum \cite{Full73,Full77,Unru76,Boul75,
sahrin}, turns out to be inequivalent to that for an inertial observer, the
familiar Minkowski vacuum. Quantum field theory in accelerated systems
contains many of special features produced by a gravitational field avoiding
some of the difficulties entailed by renormalization in a curved spacetime.
In particular, near the canonical horizon in the gravitational field, a
static spacetime may be regarded as a Rindler--like spacetime. Rindler space
is conformally related to the de Sitter space and to the Robertson--Walker
space with negative spatial curvature. As a result the expectation values of
the energy--momentum tensor for a conformally invariant field and for
corresponding conformally transformed boundaries on the de Sitter and
Robertson--Walker backgrounds can be derived from the corresponding Rindler
counterpart by the standard transformation \cite{Birrell}. Vacuum
expectation values of the energy-momentum tensor for the conformally coupled
Dirichlet and Neumann massless scalar and electromagnetic fields in four
dimensional Rindler spacetime was considered by Candelas and Deutsch \cite%
{CandD}. In Ref.\cite{setsahr} the vacuum expectation value of the surface
energy-momentum tensor is evaluated for a massless scalar field obeying a
Robin boundary condition on an infinite plane moving by uniform proper
acceleration through Fulling-Rindler vacuum. By using the conformal relation
between the Rindler and de Sitter spacetimes, in Ref.\cite{setsahr1} the
vacuum energy-momentum tensor for a scalar field is evaluated in de Sitter
spacetime in presence of a curved brane on which the field obeys the Robin
boundary condition with coordinate dependent coefficients.\newline
In this paper the vacuum expectation value of the energy-momentum tensor is
investigated for a massless conformally coupled scalar field obeying the
Dirichlet boundary condition on two curved boundaries in $k=-1$ static
Robertson--Walker backgrounds. Here we use the results of Ref.\cite{sahrin}
to generate vacuum energy---momentum tensor for the $k=-1$ Robertson--Walker
background which is conformally related to the Rindler spacetime. Previously
this method has been used in \cite{q} to derive the vacuum stress on
parallel plates for scalar field with Dirichlet boundary condition in de
Sitter spacetime. Also this method has been used in \cite{p} for the
investigation of the vacuum characteristics of the Casimir configuration on
background of conformally flat brane-world geometries for massless scalar
field with Robin boundary condition on plates.

\section{Vacuum expectation values for the energy-momentum tensor}

We will consider a conformally coupled massless scalar field $\varphi (x)$
satisfying the following equation
\begin{equation}
\left( \nabla _{\mu }\nabla ^{\mu }+\frac{1}{6}R\right) \varphi (x)=0,
\label{fieldeq}
\end{equation}%
on the background of a static $k=-1$ Robertson--Walkers spacetime. In Eq. (%
\ref{fieldeq}) $\nabla _{\mu }$ is the operator of the covariant derivative,
and $R$ is the Ricci scalar for the background spacetime. The corresponding
line element is given by the formula
\begin{equation}
ds_{{\rm RW}}^{2}=dt^{2}-\gamma dr^{2}-r^{2}(d\theta ^{2}+\sin ^{2}\theta
d\phi ^{2}),  \label{eqds}
\end{equation}%
where $\gamma =1/(1+r^{2})$. This geometry is conformally related to the
Rindler spacetime \cite{Birrell} \newline
The four-dimensional Rindler line element can be written as \cite{Birrell}
\begin{equation}
ds_{{\rm Rin}}^{2}=\xi ^{2}d\eta ^{2}-d\xi ^{2}-dy^{2}-dz^{2},\hspace{1cm}%
0<\xi <\infty  \label{rin}
\end{equation}%
Under the coordinate transformation

\begin{eqnarray}
&&\xi =\gamma ^{1/2}(1-r\gamma ^{1/2}\cos \theta )^{-1},\quad y=r\gamma
^{1/2}\sin \theta \cos \phi (1-r\gamma ^{1/2}\cos \theta )^{-1},\quad \eta
=t,  \nonumber \\
&&z=r\gamma ^{1/2}\sin \theta \sin \phi (1-r\gamma ^{1/2}\cos \theta )^{-1},
\label{coordtrans}
\end{eqnarray}%
line element (\ref{rin}) takes the form
\begin{equation}
ds_{{\rm Rin}}^{2}=\xi ^{2}[dt^{2}-\gamma dr^{2}-r^{2}(d\theta ^{2}+\sin
^{2}\theta d\phi ^{2})],  \label{rin1}
\end{equation}%
which explicitly shows the conformal relation between the Rindler and $k=-1$
Robertson--Walker space-times. Let us denote the energy--momentum tensor in
coordinates $(\eta ,\xi ,y,z)$ as $T_{\mu \nu }$, and the same tensor in
coordinates $(t,r,\theta ,\phi )$ as $T_{\mu \nu }^{\prime }$. Our main
interest in the present paper is to investigate the vacuum expectation
values (VEV's) of the energy--momentum tensor (EMT) for the field $\varphi
(x)$ in the background described by (\ref{eqds}) induced by two curved
boundaries. We will consider the case of a scalar field satisfying Dirichlet
boundary condition on the surface of the plates:
\begin{equation}
\varphi |_{\xi =\xi _{1}}=\varphi |_{\xi =\xi _{2}}=0,  \label{Dboundcond}
\end{equation}%
The presence of boundaries modifies the spectrum of the zero--point
fluctuations compared to the case without boundaries. This results in the
shift in the VEV's of the physical quantities, such as vacuum energy density
and stresses. This is the well known Casimir effect. It can be shown that
for a conformally coupled scalar by using field equation (\ref{fieldeq}) the
expression for the energy--momentum tensor can be presented in the form
\begin{equation}
T_{\mu \nu }=\nabla _{\mu }\varphi \nabla _{\nu }\varphi -\frac{1}{6}\left[
\frac{g_{\mu \nu }}{2}\nabla _{\rho }\nabla ^{\rho }+\nabla _{\mu }\nabla
_{\nu }+R_{\mu \nu }\right] \varphi ^{2},  \label{EMT1}
\end{equation}%
where $R_{\mu \nu }$ is the Ricci tensor. The quantization of a scalar filed
on background of metric Eq.(\ref{eqds}) is standard. Let $\{\varphi _{\alpha
}(x),\varphi _{\alpha }^{\ast }(x)\}$ be a complete set of orthonormalized
positive and negative frequency solutions to the field equation (\ref%
{fieldeq}), obying boundary condition (\ref{Dboundcond}). By expanding the
field operator over these eigenfunctions, using the standard commutation
rules and the definition of the vacuum state for the vacuum expectation
values of the energy-momentum tensor one obtains
\begin{equation}
\langle 0|T_{\mu \nu }(x)|0\rangle =\sum_{\alpha }T_{\mu \nu }\{\varphi {%
_{\alpha },\varphi _{\alpha }^{\ast }\}},  \label{emtvev1}
\end{equation}%
where $|0\rangle $ is the amplitude for the corresponding vacuum state, and
the bilinear form $T_{\mu \nu }\{{\varphi ,\psi \}}$ on the right is
determined by the classical energy-momentum tensor (\ref{EMT1}). Instead of
evaluating Eq. (\ref{emtvev1}) directly on background of the curved metric,
the vacuum expectation values can be obtained from the corresponding Rindler
space time results for a scalar field $\bar{\varphi}$ by using the conformal
properties of the problem under consideration. Under the conformal
transformation $g_{\mu \nu }=\Omega ^{2}\bar{g}_{\mu \nu }$ the $\bar{\varphi%
}$ field will change by the rule
\begin{equation}
\varphi (x)=\Omega ^{-1}\bar{\varphi}(x),  \label{phicontr}
\end{equation}%
where for metric (\ref{eqds}) the conformal factor is given by $\Omega =\xi
^{-1}$. In this Letter as a Rindler counterpart we will take the vacuum
energy-momentum tensor induced by a couple of infinite plates moving by
uniform proper acceleration through the Fulling-Rindler vacuum. In the
corresponding RW problem the geometry of boundaries follows from (\ref%
{coordtrans}) and is given by the equations
\begin{equation}
\frac{\gamma ^{1/2}}{1-r\gamma ^{1/2}\cos \theta }=\xi _{i},\hspace{0.5cm}%
i=1,2,  \label{curbou}
\end{equation}%
where $\xi =\xi _{i}$ are the locations of the plates in the Rindler
problem. \newline
The Casimir effect for two parallel plates moving with uniform proper
acceleration on background of the Rindler spacetime is investigated in Ref.%
\cite{sahrin} for a scalar field with Dirichlet and Neumann boundary
conditions and in \cite{sahrinR} for Robin boundary conditions. The
expectation values of the energy-momentum tensor foe a scalar field $\varphi
_{R}$ in the Fulling-Rindler vacuum can be presented in the form of the sum
\begin{equation}
<0_{R}|T_{i}^{k}|0_{R}>=<\tilde{0}_{R}|T_{i}^{k}|\tilde{0}%
_{R}>+<T_{i}^{k}>^{(b)},  \label{phiR}
\end{equation}%
where $|0_{R}>$ and $|\tilde{0}_{R}>$ are the amplitudes for the vacuum
states in the Rindler space in the presence and of absence of the plates
respectively, $<T_{i}^{k}>^{(b)}$ is the part of the vacuum energy-momentum
tensor induced by the plates. In the case of a conformally coupled massless
scalar field for the part without boundaries one has
\begin{equation}
<\tilde{0}_{R}|T_{i}^{k}|\tilde{0}_{R}>=\frac{-2\delta _{i}^{k}}{(4\pi
)^{3/2}\Gamma (3/2)\xi ^{4}}\int_{0}^{\infty }\frac{\omega
^{3}g^{(i)}(\omega )d\omega }{e^{2\pi \omega }-1},  \label{phiR1}
\end{equation}%
where the expressions for the functions $g^{(i)}(\omega )$ are presented in
Ref.\cite{Saharian1}.\newline
For a scalar field $\varphi _{R}$, satisfying the Dirichlet boundary
condition, the boundary induced part in the region between the plates has
the form \cite{sahrin}
\begin{equation}
<T_{i}^{k}>^{(b)}=<T_{i}^{k}>^{(b)}(\xi _{1},\xi )-\frac{1}{2\pi ^{2}}\delta
_{i}^{k}\int_{0}^{\infty }dkk^{3}\int_{0}^{\infty }d\omega \frac{I_{\omega
}(k\xi _{1})}{I_{\omega }(k\xi _{2})}\frac{F^{(i)}[D_{\omega }(k\xi ,k\xi
_{2})]}{D_{\omega }(k\xi _{1},k\xi _{2})},  \label{phiRr}
\end{equation}%
where
\begin{equation}
D_{\omega }(k\xi ,k\xi _{2})=I_{\omega }(k\xi _{2})K_{\omega }(k\xi
)-K_{\omega }(k\xi _{2})I_{\omega }(k\xi ),  \label{phiR11fd}
\end{equation}%
and $I_{\omega }(z)$, $K_{\omega }(z)$ are the modified Bessel functions.
The first term on the right hand-side of (\ref{phiRr}),
\begin{equation}
<T_{i}^{k}>^{(b)}(\xi _{1},\xi )=\frac{1}{2\pi ^{2}}\delta
_{i}^{k}\int_{0}^{\infty }dkk^{3}\int_{0}^{\infty }d\omega \frac{I_{\omega
}(k\xi _{1})}{K_{\omega }(k\xi _{1})}F^{(i)}[K_{\omega }(k\xi )],
\label{phiR11}
\end{equation}%
is the part induced in the region $\xi >\xi _{1}$ by the presence of a
single plane boundary located at $\xi =\xi _{1}$. The functions $%
F^{(i)}[G(z)],i=0,1,2,3$ in the expressions above are defined by the
formulae
\begin{equation}
F^{(0)}[G(z)]=\frac{1}{6}|\frac{dG(z)}{dz}|^{2}+\frac{1}{6z}\frac{d}{dz}%
|G(z)|^{2}+\frac{1}{6}(1-\frac{5\omega ^{2}}{z^{2}})|G(z)|^{2},
\label{phiR110}
\end{equation}%
\begin{equation}
F^{(1)}[G(z)]=\frac{-1}{2}|\frac{dG(z)}{dz}|^{2}-\frac{1}{6z}\frac{d}{dz}%
|G(z)|^{2}+\frac{1}{2}(1+\frac{\omega ^{2}}{z^{2}})|G(z)|^{2},
\label{phiR112}
\end{equation}%
\begin{equation}
F^{(2,3)}[G(z)]=\frac{-1}{2}|\frac{dG(z)}{dz}|^{2}+\frac{1}{6}|\frac{dG(z)}{%
dz}|^{2}+\frac{1}{6}(1+\frac{\omega ^{2}}{z^{2}})|G(z)|^{2},  \label{phiR11i}
\end{equation}%
and the indices $0,1$ correspond to the coordinates $\tau ,\xi $
respectively. The expressions in (\ref{phiRr}), (\ref{phiR11}) are given in
the coordinate system $(\eta ,\xi ,y,z)$. To find the vacuum expectation
values generated by the boundaries in the $k=-1$ static RW spacetime, first
we will consider the Rindler spacetime quantities in the coordinates $%
t,r,\theta ,\phi $ with line element (\ref{rin1}). So, it is necessary to
make the coordinate transformation to obtain $<0_{R}|\acute{T}%
_{i}^{k}|0_{R}> $. After long calculations for the part induced by the
boundaries we find the following expressions
\begin{equation}
<\acute{T}_{t}^{t}>^{(b)}=<T_{\eta }^{\eta }>^{(b)}.  \label{tt}
\end{equation}

\begin{eqnarray}
<\acute{T}_{\theta }^{\theta }>^{(b)} &=&\Bigg\{\frac{b^{2}}{\cos ^{2}\phi }+%
\frac{r^{2}\gamma ^{2}\sin ^{2}\theta }{a^{2}}(1-r\gamma ^{1/2}\cos \theta
)^{-4}\Big[\gamma ^{1/2}\sin \theta \cos \phi (1-r\gamma ^{1/2}\cos \theta
)^{-1}  \nonumber \\
&&+ra\sin \theta \cos \phi \Big]^{2}(\cot ^{2}\phi -\tan ^{2}\phi )\Bigg\}%
^{-1}\Big[(1-\cot ^{2}\phi )<T_{y}^{y}>^{(b)}  \nonumber \\
&&+\left( \frac{\gamma ^{1/2}\sin \theta \cos \phi }{1-r\gamma ^{1/2}\cos
\theta }+ra\sin \theta \cos \phi \right) \frac{\cot ^{2}\phi -\tan ^{2}\phi
}{a^{2}}<T_{\xi }^{\xi }>^{(b)}\Big]  \label{tet}
\end{eqnarray}%
\begin{equation}
<\acute{T}_{r}^{r}>^{(b)}=\frac{1}{a^{2}}\left[ <T_{\xi }^{\xi
}>^{(b)}-r^{2}\gamma ^{2}(1-r\gamma ^{1/2}\cos \theta )^{-4}\sin ^{2}\theta <%
\acute{T}_{\theta }^{\theta }>^{(b)}\right] .  \label{rr}
\end{equation}%
\begin{eqnarray}
<\acute{T}_{\phi }^{\phi }>^{(b)} &=&\frac{(1-r\gamma ^{1/2}\cos \theta
)^{-2}}{r^{2}\gamma \sin ^{2}\theta \cos ^{2}\phi }\Bigg\{<T_{z}^{z}>^{(b)}-%
\Big[\gamma ^{1/2}\sin \theta \sin \phi (1-r\gamma ^{1/2}\cos \theta )^{-1}
\nonumber \\
&&+ra\sin \theta \sin \phi \Big]^{2}<\acute{T}_{r}^{r}>^{(b)}-\Big[r\gamma
^{1/2}\cos \theta \sin \phi (1-r\gamma ^{1/2}\cos \theta )^{-1}  \nonumber \\
&&-r^{2}\gamma \sin ^{2}\theta \sin \phi (1-r\gamma ^{1/2}\cos \theta )^{-2}%
\Big]^{2}<\acute{T}_{\theta }^{\theta }>^{(b)}\Bigg\}  \label{phi}
\end{eqnarray}%
where
\begin{equation}
a=\gamma (1-r\gamma ^{1/2}\cos \theta )^{-1}\Big[r\gamma ^{1/2}+\cos \theta
(1-r^{2}\gamma )(1-r\gamma ^{1/2}\cos \theta )^{-1}\Big].  \label{a}
\end{equation}%
\begin{equation}
b=r\gamma ^{1/2}\cos \phi (1-r\gamma ^{1/2}\cos \theta )^{-1}\Big[\cos
\theta -r\gamma ^{1/2}(1-r\gamma ^{1/2}\cos \theta )^{-1}\Big].  \label{b}
\end{equation}%
Therefore, we can write the RW vacuum expectation values in the form
similar to (\ref{phiR}):
\begin{equation}
<0_{RW}|T_{i}^{k}|0_{RW}>=<\tilde{0}_{RW}|T_{i}^{k}|\tilde{0}%
_{RW}>+<T_{i}^{k}>_{RW}^{(b)},  \label{phiR10}
\end{equation}%
where $<\tilde{0}_{RW}|T_{i}^{k}|\tilde{0}_{RW}>$ are the vacuum expectation
values in $k=-1$ RW space without boundaries and the part $<T_{i}^{k}>^{(b)}$
is induced by the plates. Conformally transforming the Rindler results one
finds
\begin{equation}
<\tilde{0}_{RW}|T_{i}^{k}|\tilde{0}_{RW}>=\Omega ^{-4}<\tilde{0}%
_{R}|T_{i}^{k}|\tilde{0}_{R}>+<T_{i}^{k}>^{(an)},  \label{phiR110}
\end{equation}%
\begin{equation}
<T_{i}^{k}>_{RW}^{(b)}=\Omega ^{-4}<\acute{T}_{i}^{k}>^{(b)},
\label{phiR111}
\end{equation}%
The second term in Eq.(\ref{phiR110}) can be rewritten in the form
\begin{equation}
<T_{i}^{k}>^{(an)}=-\frac{1}{2880}[\frac{1}{6}\tilde{H}_{\nu }^{(1)\mu }-%
\tilde{H}_{\nu }^{(3)\mu }]  \label{frw}
\end{equation}%
Fortunately for the static RW space-time with $k=-1$ the boundary free part
is zero \cite{Birrell}
\begin{equation}
<\tilde{0}_{RW}|T_{i}^{k}|\tilde{0}_{RW}>=0.  \label{freepa}
\end{equation}%
Therefore the vacuum energy-momentum tensor can be evaluated exactly as:
\begin{equation}
<0_{RW}|T_{i}^{k}|0_{RW}>=\Omega ^{-4}<\acute{T}_{i}^{k}>^{(b)},
\label{phiR111}
\end{equation}%
where non-zero components of $<\acute{T}_{i}^{k}>^{(b)}$ are given by Eqs.(%
\ref{tt})-(\ref{phi}).

\section{Conclusion}

In the present paper we have investigated the Casimir effect for a
conformally coupled massless scalar field between two courved
boundaries, on background of the static $k=-1$ Robertson--Walker
spacetime. We have assumed that the scalar field satisfies Dirichlet
boundary condition on the curved boundaries. The $k=-1$
Robertson--Walker spacetime is conformally related to the Rindler
spacetime, then the vacuum expectation values of the energy-momentum
tensor are derived from the corresponding Rindler spacetime results
by using the conformal properties of the problem. The vacuum
expectation value of the energy-momentum tensor for the curved
boundaries in Robertson--Walker spacetime consists of two parts
given in Eq.(\ref{phiR10}). The first one corresponds to the purely
RW contribution when the boundary is absent. For the static RW
space-time with $k=-1$ this boundary free part is zero
\cite{Birrell}. The second part in the vacuum energy-momentum tensor
is due to the imposition of boundary conditions on the fluctuating
quantum field. The corresponding components are related to the
vacuum energy-momentum tensor in the Rindler spacetime by Eq.
(\ref{phiR111}) and the Rindler tensor is given by equations
(\ref{tt})-(\ref{phi}).

\vspace{3mm}

\section{Acknowledgment}

The author are indebted to the referee for his comments that improved the
paper drastically.

\end{document}